%% file: main.tex
\pdfoutput=1
\documentclass[12pt,a4paper]{article}
\usepackage{ifthen} 
\newboolean{pdflatex}
\setboolean{pdflatex}{true} 

\newboolean{articletitles}
\setboolean{articletitles}{true} 

\newboolean{uprightparticles}
\setboolean{uprightparticles}{false} 

\def\paperauthors{LHCb collaboration} 
\def\paperasciititle{Evidence for the rare decay B+ -> Lam_bar p mu+ mu-} 
\def\papertitle{Evidence for the rare decay \btlpmm} 
\def\paperkeywords{{High Energy Physics}, {LHCb}} 
\def\papercopyright{\the\year\ CERN for the benefit of the LHCb collaboration} 
\def\paperlicence{CC BY 4.0 licence}
\def\paperlicenceurl{https://creativecommons.org/licenses/by/4.0/}

\newif\ifEnableSectionTOCLinks
\EnableSectionTOCLinksfalse 

\input{preamble}
\usepackage{longtable} 

\begin{document}

\renewcommand{\thefootnote}{\fnsymbol{footnote}}
\setcounter{footnote}{1}

\input{title-LHCb-PAPER}

\renewcommand{\thefootnote}{\arabic{footnote}}
\setcounter{footnote}{0}

\cleardoublepage

\pagestyle{plain} 
\setcounter{page}{1}
\pagenumbering{arabic}


\input{body}

\input{acknowledgements}

\clearpage

\addcontentsline{toc}{section}{References}
\bibliographystyle{LHCb}
\bibliography{main,standard,LHCb-PAPER,LHCb-CONF,LHCb-DP,LHCb-TDR}
 
\begingroup
  \makeatletter
    \let\addcontentsline\@gobblethree
  \makeatother
  \input{endmatter}  
\endgroup

\clearpage
\newpage
\input{Authorship_LHCb-PAPER-2025-051}
\end{document}

%% file: preamble.tex
\usepackage[top=1in, bottom=1.25in, left=1in, right=1in]{geometry}

\usepackage{microtype}
\usepackage{lineno}  
\usepackage{xspace} 
\usepackage{caption} 

\usepackage{graphicx}  
\usepackage{color}
\usepackage{colortbl}
\graphicspath{{./figs/}} 

\usepackage{amsmath} 
\usepackage{amssymb}
\usepackage{amsfonts}
\usepackage{upgreek} 

\newcommand*\patchAmsMathEnvironmentForLineno[1]{%
\expandafter\let\csname old#1\expandafter\endcsname\csname #1\endcsname
\expandafter\let\csname oldend#1\expandafter\endcsname\csname
end#1\endcsname
 \renewenvironment{#1}%
   {\linenomath\csname old#1\endcsname}%
   {\csname oldend#1\endcsname\endlinenomath}%
}
\newcommand*\patchBothAmsMathEnvironmentsForLineno[1]{%
  \patchAmsMathEnvironmentForLineno{#1}%
  \patchAmsMathEnvironmentForLineno{#1*}%
}
\AtBeginDocument{%
\patchBothAmsMathEnvironmentsForLineno{equation}%
\patchBothAmsMathEnvironmentsForLineno{align}%
\patchBothAmsMathEnvironmentsForLineno{flalign}%
\patchBothAmsMathEnvironmentsForLineno{alignat}%
\patchBothAmsMathEnvironmentsForLineno{gather}%
\patchBothAmsMathEnvironmentsForLineno{multline}%
\patchBothAmsMathEnvironmentsForLineno{eqnarray}%
}

\usepackage[pdftex,
            pdfauthor={\paperauthors},
            pdftitle={\paperasciititle},
            pdfkeywords={\paperkeywords}]{hyperref}
\usepackage{hyperxmp}
\hypersetup{
    pdfcopyright={Copyright (C) \papercopyright},
    pdflicenseurl={\paperlicenceurl}
}

\usepackage[colorinlistoftodos,textsize=scriptsize]{todonotes}

\usepackage[bottom,flushmargin,hang,multiple]{footmisc}

\usepackage[all]{hypcap} 

\input{lhcb-symbols-def} 

\input{our-symbols-def}  

\hypersetup{
  colorlinks   = true, 
  urlcolor     = blue, 
  linkcolor    = blue, 
  citecolor    = red   
}

\ifEnableSectionTOCLinks
    \usepackage[explicit]{titlesec} 
    
    \let\oldcontentsline\contentsline
    \renewcommand

    \titleformat{\section}{\normalfont\Large\bf}{\hyperlink{tocsection.\thesection}{{\thesection} \parbox[t]{\dimexpr\textwidth-1pc}{#1}}}{1pc}{}

    \titleformat{\subsection}{\normalfont\bf}{\hyperlink{tocsubsection.\thesubsection}{{\thesubsection} \parbox[t]{\dimexpr\textwidth-1pc}{#1}}}{1pc}{}

    \titleformat{name=\section,numberless}[display]{}{}{0pt}{\normalfont\Huge\bfseries #1}
\fi

\usepackage{cite} 
\usepackage{mciteplus}

%% file: lhcb-symbols-def.tex
\usepackage{xspace}
\usepackage{upgreek}

\def\lhcb   {\mbox{LHCb}\xspace}

\def\babar  {\mbox{BaBar}\xspace}
\def\belle  {\mbox{Belle}\xspace}

\def\MagUp {\mbox{\em Mag\kern -0.05em Up}\xspace}

\ifthenelse{\boolean{uprightparticles}}%
{

 \def\Pmu         {\ensuremath{\upmu}\xspace}

 \def\Ppi         {\ensuremath{\uppi}\xspace}

 \def\Ppsi        {\ensuremath{\uppsi}\xspace}

 \def\PDelta      {\ensuremath{\Delta}\xspace}
 \def\PXi         {\ensuremath{\Xi}\xspace}
 \def\PLambda     {\ensuremath{\Lambda}\xspace}
 \def\PSigma      {\ensuremath{\Sigma}\xspace}
 \def\POmega      {\ensuremath{\Omega}\xspace}
 \def\PUpsilon    {\ensuremath{\Upsilon}\xspace}
 \let\oldPi\Pi
 \def\PPi         {\ensuremath{\oldPi}\xspace}

 \def\PB      {\ensuremath{\mathrm{B}}\xspace}
 \def\PD      {\ensuremath{\mathrm{D}}\xspace}
 \def\PJ      {\ensuremath{\mathrm{J}}\xspace}
 \def\PK      {\ensuremath{\mathrm{K}}\xspace}
 \def\Pb      {\ensuremath{\mathrm{b}}\xspace}
 \def\Pc      {\ensuremath{\mathrm{c}}\xspace}

 \def\Ps      {\ensuremath{\mathrm{s}}\xspace}

 \def\thebaroffset{0.0em}
}
{

 \def\Pmu         {\ensuremath{\mu}\xspace}

 \def\Ppi         {\ensuremath{\pi}\xspace}

 \def\Ppsi        {\ensuremath{\psi}\xspace}
 
 \mathchardef\PDelta="7101
 \mathchardef\PXi="7104
 \mathchardef\PLambda="7103
 \mathchardef\PSigma="7106
 \mathchardef\POmega="710A
 \mathchardef\PUpsilon="7107
 \mathchardef\PPi="7105
 \def\PB      {\ensuremath{B}\xspace}
 \def\PD      {\ensuremath{D}\xspace}
 \def\PJ      {\ensuremath{J}\xspace}
 \def\PK      {\ensuremath{K}\xspace}
 \def\Pb      {\ensuremath{b}\xspace}
 \def\Pc      {\ensuremath{c}\xspace}

 \def\Ps      {\ensuremath{s}\xspace}

 \def\thebaroffset{0.18em}
}
\newcommand{\offsetoverline}[2][\thebaroffset]{\kern #1\overline{\kern -#1 #2}}%

\makeatletter
\ifcase \@ptsize \relax
  \newcommand{\miniscule}{\@setfontsize\miniscule{4}{5}}
\or
  \newcommand{\miniscule}{\@setfontsize\miniscule{5}{6}}
\or
  \newcommand{\miniscule}{\@setfontsize\miniscule{5}{6}}
\fi
\makeatother

\DeclareRobustCommand{\optbar}[1]{\shortstack{{\miniscule (\rule[.5ex]{1.25em}{.18mm})}
  \\ [-.7ex] $#1$}}

\def\mumu       {{\ensuremath{\Pmu^+\Pmu^-}}\xspace}

\def\squark    {{\ensuremath{\Ps}}\xspace}

\def\cquark    {{\ensuremath{\Pc}}\xspace}

\def\bquark    {{\ensuremath{\Pb}}\xspace}

\def\pion   {{\ensuremath{\Ppi}}\xspace}

\def\pip    {{\ensuremath{\pion^+}}\xspace}

\def\kaon    {{\ensuremath{\PK}}\xspace}

\def\KorKbar {\kern \thebaroffset\optbar{\kern -\thebaroffset \PK}{}\xspace}

\def\Kp      {{\ensuremath{\kaon^+}}\xspace}

\def\D       {{\ensuremath{\PD}}\xspace}

\def\DorDbar {\kern \thebaroffset\optbar{\kern -\thebaroffset \PD}\xspace}

\def\Dp      {{\ensuremath{\D^+}}\xspace}
\def\Dm      {{\ensuremath{\D^-}}\xspace}

\def\DpDm    {\ensuremath{\Dp {\kern -0.16em \Dm}}\xspace}

\def\B       {{\ensuremath{\PB}}\xspace}

\def\BorBbar {\kern \thebaroffset\optbar{\kern -\thebaroffset \PB}\xspace}

\def\Bd      {{\ensuremath{\B^0}}\xspace}

\def\BdorBdbar {\kern \thebaroffset\optbar{\kern -\thebaroffset \Bd}\xspace}
\def\Bu      {{\ensuremath{\B^+}}\xspace}

\def\Bp      {{\ensuremath{\Bu}}\xspace}

\def\Bs      {{\ensuremath{\B^0_\squark}}\xspace}

\def\BsorBsbar {\kern \thebaroffset\optbar{\kern -\thebaroffset \Bs}\xspace}

\def\jpsi     {{\ensuremath{{\PJ\mskip -3mu/\mskip -2mu\Ppsi}}}\xspace}

\def\Y#1S{\ensuremath{\PUpsilon{(#1S)}}\xspace}

\def\Lbar        {{\ensuremath{\offsetoverline{\PLambda}}}\xspace}
\def\LorLbar     {\kern \thebaroffset\optbar{\kern -\thebaroffset \PLambda}\xspace}

\def\BF         {{\ensuremath{\mathcal{B}}}\xspace}

\newcommand{\decay}[2]{\mbox{\ensuremath{#1\!\to #2}}\xspace}

\def\to                 {\ensuremath{\rightarrow}\xspace}

\def\AT#1     {\ensuremath{A_{\mathrm{T}}^{#1}}\xspace}           

\def\C#1      {\ensuremath{\mathcal{C}_{#1}}\xspace}                       
\def\Cp#1     {\ensuremath{\mathcal{C}_{#1}^{'}}\xspace}                    
\def\Ceff#1   {\ensuremath{\mathcal{C}_{#1}^{\mathrm{(eff)}}}\xspace}        
\def\Cpeff#1  {\ensuremath{\mathcal{C}_{#1}^{'\mathrm{(eff)}}}\xspace}       
\def\Ope#1    {\ensuremath{\mathcal{O}_{#1}}\xspace}                       
\def\Opep#1   {\ensuremath{\mathcal{O}_{#1}^{'}}\xspace}                    

\newcommand{\aunit}[1]{\ensuremath{\text{\,#1}}}

\newcommand{\tev}{\aunit{Te\kern -0.1em V}\xspace}
\newcommand{\gev}{\aunit{Ge\kern -0.1em V}\xspace}
\newcommand{\mev}{\aunit{Me\kern -0.1em V}\xspace}
\newcommand{\kev}{\aunit{ke\kern -0.1em V}\xspace}
\newcommand{\ev}{\aunit{e\kern -0.1em V}\xspace}

\newcommand{\mevc}{\ensuremath{\aunit{Me\kern -0.1em V\!/}c}\xspace}
\newcommand{\gevc}{\ensuremath{\aunit{Ge\kern -0.1em V\!/}c}\xspace}
\newcommand{\mevcc}{\ensuremath{\aunit{Me\kern -0.1em V\!/}c^2}\xspace}
\newcommand{\gevcc}{\ensuremath{\aunit{Ge\kern -0.1em V\!/}c^2}\xspace}

\def\fb   {\ensuremath{\aunit{fb}}\xspace}
\def\invfb   {\ensuremath{\fb^{-1}}\xspace}

\newcommand{\chisq}{\ensuremath{\chi^2}\xspace}

\newcommand{\chisqip}{\ensuremath{\chi^2_{\text{IP}}}\xspace}

\def\gsim{{~\raise.15em\hbox{$>$}\kern-.85em
          \lower.35em\hbox{$\sim$}~}\xspace}
\def\lsim{{~\raise.15em\hbox{$<$}\kern-.85em
          \lower.35em\hbox{$\sim$}~}\xspace}

\def\pt         {\ensuremath{p_{\mathrm{T}}}\xspace}

\def\evtgen     {\mbox{\textsc{EvtGen}}\xspace}

\def\geant      {\mbox{\textsc{Geant4}}\xspace}

\def\photos     {\mbox{\textsc{Photos}}\xspace}

\def\pythia     {\mbox{\textsc{Pythia}}\xspace}

\def\tell1  {TELL1\xspace}
\def\ukl1   {UKL1\xspace}

\newcommand{\lhcborcid}[1]{\href{https://orcid.org/#1}{\hspace*{0.1em}\raisebox{-0.45ex}{\includegraphics[width=1em]{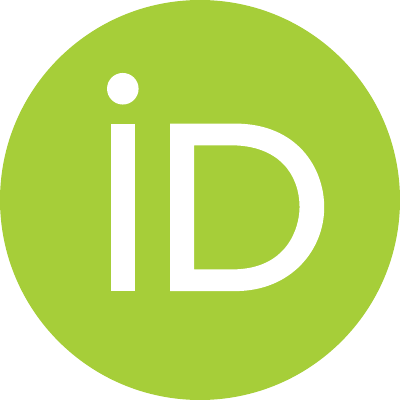}}}}


%% file: our-symbols-def.tex
\newcommand{\lamb}{\ensuremath{\Lbar}\xspace}

\newcommand{\lamp}{\ensuremath{\Lbar p}\xspace}

\newcommand{\more}{\ensuremath{>}\xspace}

\newcommand{\btsll}{\decay{b}{s\ell^+ \ell^-}} 

\newcommand{\btlpmm}{\decay{B^{+}}{\Lbar p \mu^{+} \mu^{-}}} 

\newcommand{\btjpsilp}{\decay{B^{+}}{\jpsi \Lbar p}}

\newcommand{\btjpsimmlp}{\decay{B^{+}}{\jpsi(\to \mu^{+} \mu^{-}) \Lbar p}} 

\newcommand{\btlppipi}{\decay{B^{+}}{\Lbar p \pi^{+} \pi^{-}}}

\newcommand{\btlpkk}{\decay{B^{+}}{\Lbar p K^{+} K^{-}}}

%% file: title-LHCb-PAPER.tex
\begin{titlepage}
\pagenumbering{roman}

\vspace*{-1.5cm}
\centerline{\large EUROPEAN ORGANIZATION FOR NUCLEAR RESEARCH (CERN)}
\vspace*{1.5cm}
\noindent
\begin{tabular*}{\linewidth}{lc@{\extracolsep{\fill}}r@{\extracolsep{0pt}}}
\ifthenelse{\boolean{pdflatex}}
{\vspace*{-1.5cm}\mbox{\!\!\!\includegraphics[width=.14\textwidth]{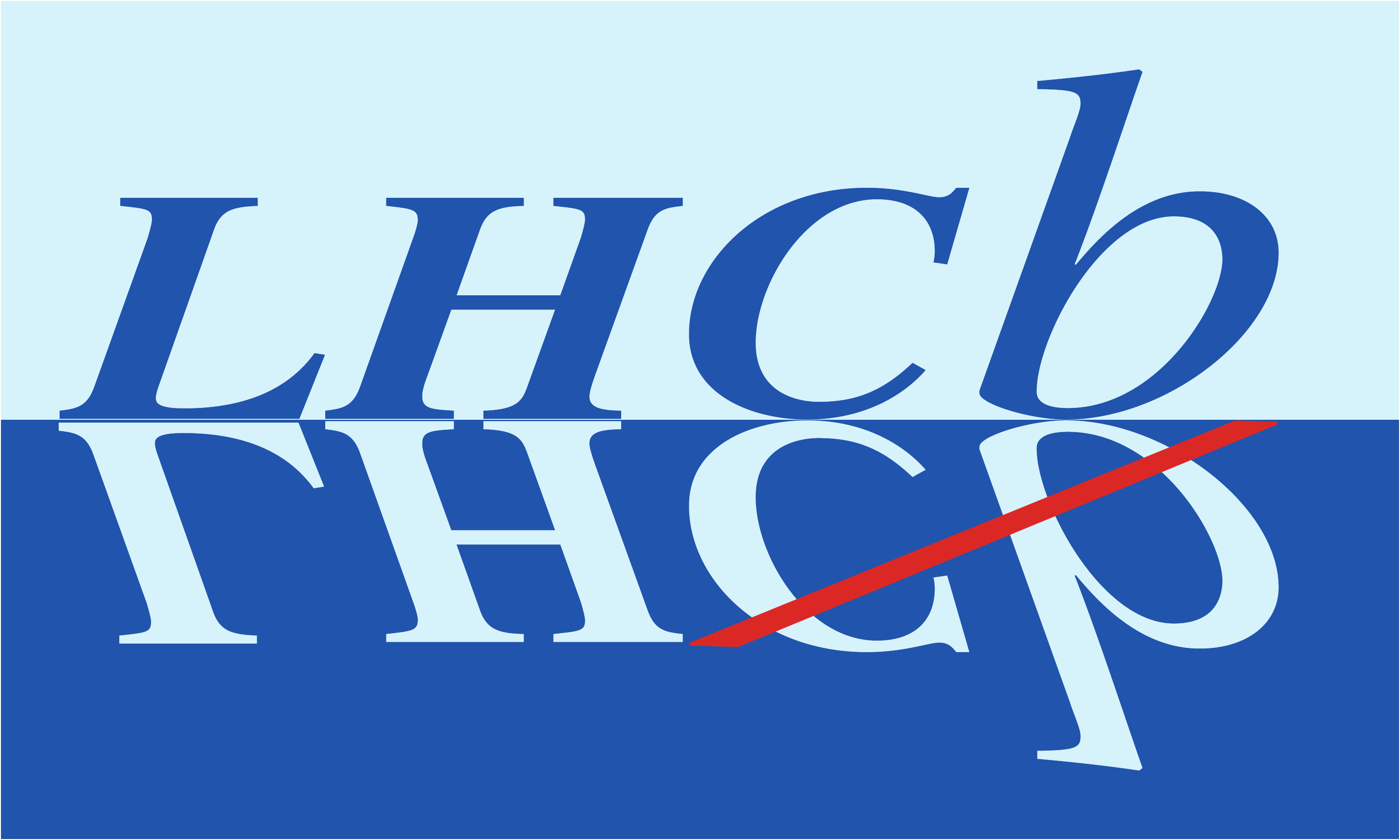}} & &}%
{\vspace*{-1.2cm}\mbox{\!\!\!\includegraphics[width=.12\textwidth]{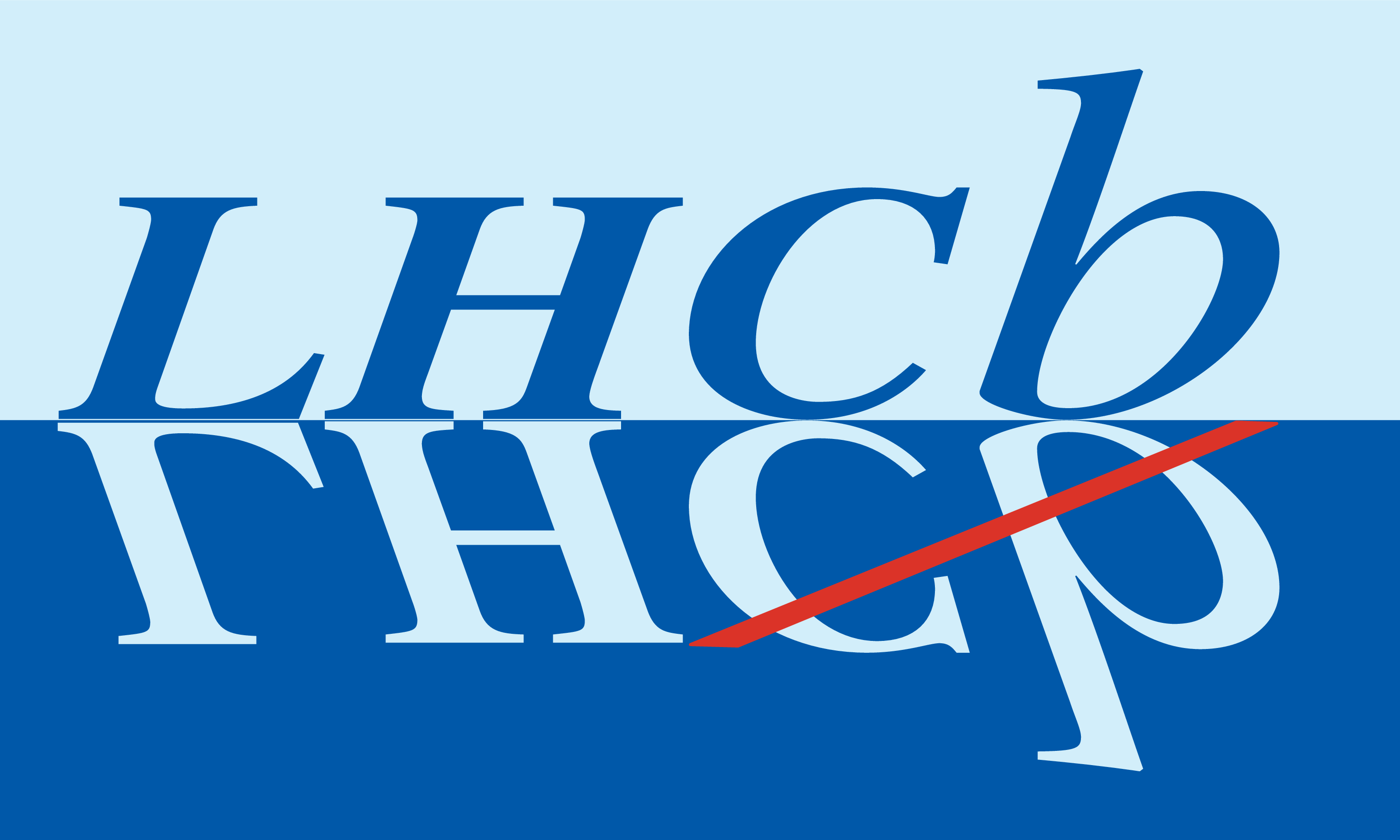}} & &}%
\\
 & & CERN-EP-2025-277 \\  
 & & LHCb-PAPER-2025-051 \\  
 & & July 29, 2026 \\
 & & \\
\end{tabular*}

\vspace*{4.0cm}

{\normalfont\bfseries\boldmath\huge
\begin{center}
  \papertitle 
\end{center}
}

\vspace*{2.0cm}

\begin{center}
\paperauthors\footnote{Authors are listed at the end of this paper.}
\end{center}

\vspace{\fill}

\begin{abstract}
  \noindent
  A search for the rare decay $\decay{B^{+}}{\Lbar p \mu^{+} \mu^{-}}$ is performed using proton-proton collision data recorded by the $\lhcb$ experiment at a center-of-mass energy of $\sqrt{s}= 13\tev$, corresponding to an integrated luminosity of 5.4$\invfb$. An excess of events is found with respect to the background-only expectation, with a signal significance of 3.5 standard deviations, in the 
  low invariant-mass region of $m(\Lbar p)<2.8\gevcc$. The branching fraction is measured to be 
  \mbox{${\cal B}_{\rm low}(\decay{B^{+}}{\Lbar p \mu^{+} \mu^{-}})=\left(1.70 ^{+0.65}_{-0.56}(\rm stat) \pm 0.17(\rm syst) \pm 0.14(\rm ext)\right) \times 10^{-8}$},   
 where the last uncertainty is due to external inputs on \mbox{${\cal B}(\decay{\Bp}{\jpsi \Lbar p})\times {\cal B}(\decay{\jpsi}{\mu^+\mu^-})$}. With no significant signal observed in the high $m(\Lbar p)$ region above 2.8$\gevcc$, an upper limit is set to be  
  ${\cal B}_{\rm high}(\decay{B^{+}}{\Lbar p \mu^{+} \mu^{-}})<2.8\,(3.7) \times 10^{-9}$
  at the 90\% (95\%) confidence level. 
  
\end{abstract}

\vspace*{2.0cm}

\begin{center}
    Published in
    Phys.~Rev.~Lett. {\bf 137} (2026) 051801
\end{center}

\vspace{\fill}

{\footnotesize 
\centerline{\copyright~\papercopyright. \href{\paperlicenceurl}{\paperlicence}.}}
\vspace*{2mm}

\end{titlepage}

\newpage
\setcounter{page}{2}
\mbox{~}

%% file: body.tex
Rare decays of heavy-flavored hadrons, particularly those mediated by flavor-changing neutral currents (FCNCs), serve as powerful probes of the Standard Model (SM) and potential signatures of New Physics (NP). In the SM, FCNC transitions such as \btsll occur only via loop-level processes and are thus highly suppressed. This suppression makes rare decays especially sensitive to contributions from non-SM particles, including new gauge bosons, leptoquarks, or supersymmetric states~\cite{Buchalla:1996vs,Buras:2003jf,Heeck:2014zfa,Langacker:2008yv,Buchmuller:1986zs,Becirevic:2017jtw,Gabbiani:1996hi,Buras:2000dm}. Among these, the decay \btlpmm is of particular interest. It combines the rare \btsll transition with a baryonic final state, which is  comparatively less explored than mesonic channels. Baryonic decays provide a complementary environment to mesonic modes for studying nonperturbative hadronization dynamics, with their rich multibody kinematics and angular observables offering sensitive probes of potential NP effects~\cite{Suzuki:2002cp}. Theoretical predictions are limited by nonperturbative effects, making these decays a valuable testing ground for the SM, with experimental measurements providing essential constraints. Theoretical studies of this channel are scarce. Currently, only one SM prediction of the branching fraction exists, ${\cal B}(\btlpmm) = (1.08^{+0.82}_{-0.51}) \times 10^{-7}$~\cite{Dutta:2014yta}, based on the baryonic form factors of Ref.~\cite{Geng:2012qn}, where the related mode \decay{\Bp}{\lamp \nu \bar \nu} was studied. A more recent SM prediction for ${\cal B}(\decay{\Bp}{\lamp \nu \bar \nu})$ reports $(3.5 \pm 1.0)\times 10^{-8}$~\cite{Hsiao:2022uzx}, nearly an order of magnitude smaller than the earlier estimate~\cite{Geng:2012qn}, owing to an updated set of baryonic form factors. This implies that a substantially lower prediction for ${\cal B}(\btlpmm)$ is expected when using the form factors of Ref.~\cite{Hsiao:2022uzx}. Experimentally, the \babar collaboration has set an upper limit of ${\cal B}(\decay{\Bp}{\lamp\nu\bar\nu}) < 3.0 \times 10^{-5}$ at the 90\% confidence level~\cite{BaBar:2019awu}, while no measurement exists for $\btlpmm$.

Furthermore, the decay \btlpmm\ offers unique opportunities to study baryonic effects in heavy-flavor decays. In particular, a pronounced enhancement near the \lamp threshold is expected in the baryon-antibaryon invariant-mass distribution. This behavior is consistent with the so-called threshold enhancement mechanism~\cite{Hou:2000bz,BaBar:2014omp}, which has been widely observed in other baryonic $B$ decays~\cite{Belle:2003pwf,LHCb-PAPER-2017-005}. Previous measurements by the \belle collaboration of related modes such as \btlpkk~\cite{Belle:2018ies} and \btlppipi~\cite{Belle:2009ixu}, as well as the recent observation of \decay{\Bp}{\lamp \bar p p} by the LHCb collaboration~\cite{LHCb-PAPER-2025-032}, have confirmed this phenomenon. In addition, the multibody structure of \btlpmm\ provides clean access to possible intermediate resonances, including excited kaon states~\cite{CMS:2019kbn} or exotic structures such as $X(2085)$~\cite{BESIII:2023kgz}, thereby further enriching its discovery potential.

This Letter presents the first search for the rare decay \btlpmm, with \decay{\Lbar}{ \bar p\pip}, using proton-proton ($pp$) collision data collected by the \lhcb experiment over the period 2016--2018 at a center-of-mass energy of $\sqrt{s}=13\tev$ and corresponding to an integrated luminosity of approximately 5.4\invfb. 
The inclusion of charge-conjugate processes is implied throughout this Letter. 
The branching fraction of the signal decay is measured relative to that of the topologically similar normalization mode, \btjpsilp, with \decay{\jpsi}{\mumu} and \decay{\Lbar}{ \bar p\pip}, using a simultaneous fit across all data-taking periods.

The \lhcb detector~\cite{LHCb-DP-2008-001,LHCb-DP-2014-002} is a single-arm forward spectrometer covering the pseudorapidity range $2<\eta<5$, designed for the study of particles containing \bquark or \cquark quarks. The detector used in this analysis includes a high-precision tracking system consisting of a silicon-strip vertex detector (VELO) surrounding the $pp$ interaction region~\cite{LHCb-DP-2014-001}, a large-area silicon-strip detector located upstream of a dipole magnet with a bending power of about $4{\mathrm{\,T\,m}}$, and three stations of silicon-strip detectors and straw drift tubes~\cite{LHCb-DP-2017-001} placed downstream of the magnet. The tracking system provides measurements of the track momentum and impact parameter (IP) and is used to reconstruct primary vertices (PVs). 
Different types of charged hadrons are distinguished using particle identification (PID) information from two ring-imaging Cherenkov detectors~\cite{LHCb-DP-2012-003}.  Muons are identified by a system comprising alternating layers of iron and multiwire proportional chambers. 
 The \decay{\Lbar}{ \bar p\pip} decay is reconstructed in two distinct categories depending on whether the proton and pion tracks are reconstructed in all tracking detectors (\emph{long}) or in all but the VELO (\emph{downstream}). The $\Lbar$ decay products must have an invariant mass between 1105 and 1130\mevcc. %
 While candidates in the \emph{long} category have better mass, momentum, and vertex resolution than those in the \emph{downstream} category, the \emph{downstream} category accounts for approximately two‐thirds of the total $\Lbar$ yield.

The online event selection is performed by a trigger~\cite{LHCb-DP-2012-004}, which consists of a hardware stage, based on information from the calorimeter and muon systems, followed by a software stage, which applies a full event reconstruction. The trigger strategy follows the standard two-level approach used in
previous LHCb analyses with dimuon final states, {\it e.g.}, $B_s^0 \to \mu^+ \mu^-$~\cite{LHCb-PAPER-2021-008}. Triggered data further undergo a centralized, offline processing step to deliver physics-analysis-ready data across the entire \lhcb physics program~\cite{Stripping}.

Simulated samples are used to study the properties of the signal, normalization, and background channels. 
The generation of $pp$ collisions is performed using \pythia~\cite{Sjostrand:2007gs}, with a specific \lhcb configuration~\cite{LHCb-PROC-2010-056}. Decays of unstable particles are described by \evtgen~\cite{Lange:2001uf}, in which final-state radiation is generated using \photos~\cite{davidson2015photos}.
The interactions of the generated particles with the detector material and their responses, are implemented using the \geant toolkit~\cite{Agostinelli:2002hh,Allison:2006ve}, as described in Ref.~\cite{LHCb-PROC-2011-006}. %
Signal and normalization decays are generated according to a phase-space decay model.

Offline candidates are selected exploiting the characteristic topology and kinematic features of four-body decays to a final state consisting of a $\lamb$ baryon, a proton, and a pair of muons of opposite charge. 
Candidates are classified as \btjpsilp when the dimuon invariant mass lies within 50\mevcc of the known \jpsi mass~\cite{PDG2024} and as \btlpmm when it is below 2.85\gevcc, to veto the charmonium contribution. 
All charged tracks except those used to reconstruct $\Lbar$ candidates must originate from within the VELO and are required to have a good track-fit quality. Both the \btlpmm and the \btjpsilp~decay chains are fitted constraining the mass of the \Lbar candidate to its known value~\cite{PDG2024}. The PV most consistent with the \Bp candidate’s flight direction is taken as its associated PV. This assignment is obtained by requiring the cosine of the angle between the \Bp momentum and the displacement vector from the associated PV to the decay vertex to exceed 0.999. The \Bp candidate is further required to have a large decay-length %
$\chisq$, defined as the increase in vertex-fit $\chisq$ when the decay vertex is constrained to the PV, 
ensuring a well-resolved displacement.

To further reduce the combinatorial background, a multivariate classifier is devised based on a
 boosted decision tree (BDT) classifier~\cite{Breiman,AdaBoost} implemented in the TMVA toolkit~\cite{Hocker:2007ht,*TMVA4}. This classifier combines different kinematic and geometric variables. For each $\lamb$ category, a BDT is trained using the signal simulation sample and data sidebands, defined as candidates falling in the mass range $m(\lamp \mumu)\more 5400$\mevcc. 
The BDT combines several kinematic and geometric variables of the \Bp, the \lamb, and the final-state particles.
For the \Bp candidates, the variables used are: the transverse momentum (\pt), the decay-length significance (DLS), the angle between the momentum and the flight direction, the $\chi^{2}_{\mathrm{IP}}$ (the change in the PV vertex-fit $\chi^{2}$ when the particle is included or removed), and the $\chi^{2}$ of the kinematic fit.
For the final-state particles, the BDT uses the sum of the $\chi^{2}_{\mathrm{IP}}$ values and, for \emph{downstream} candidates only, the sum of their pseudorapidities.
For the \lamb candidates, the variables considered are the angle between the momentum and the flight direction. In addition, the decay-length $\chi^{2}$ and the $\lamb$ DLS %
are included for \emph{long} candidates, while the $\chi^{2}_{\mathrm{IP}}$ is used only for \emph{downstream} candidates.

The final selection is based on the BDT output, and PID information of the final-state muons and protons. The selection criteria on these variables are optimized on a three-dimensional grid to maximize the Punzi figure of merit, defined as ${\varepsilon}/({\frac{a}{2}+\sqrt{N_B}})$~\cite{Punzi:2003bu}, where  $\varepsilon$ is the signal efficiency, $a=5$ specifies the target significance of the signal component, and $N_B$ is the expected background yield within the signal region defined as $| m(\lamb p\mumu)-m_{B^+}|<20\mevcc$. Here $m_{B^+}$ is the known mass of the $B^+$ meson~\cite{PDG2024}.
The BDT is trained on the $\btlpmm$ decay and is also applied to the $\btjpsilp$ normalization mode from the same \emph{long} or \emph{downstream} category, maintaining the same requirements on the multivariate output, as well as on all PID variables.

The reconstruction efficiencies for both the signal and normalization channels are evaluated using the simulated decays. %
The simulated samples are corrected to account for known data-simulation differences in the track reconstruction, PID, and trigger efficiencies using calibration samples. Moreover, to correct the kinematic distributions in the simulated samples for both the signal and
normalization channels, each candidate is assigned a weight derived from background-subtracted \mbox{$\btjpsilp$} candidates. The weighting is based on the $\pt$ and $\chisqip$ of the \Bp candidate, the $\chi^2$ of its kinematic fit, the cosine of the angle between the \Bp candidate momentum vector and the displacement vector connecting the associated PV to the \Bp decay vertex, and the track multiplicity in the event. %
In addition, for the normalization channel, the $m(\jpsi\lamb)$ versus $m(\jpsi p)$ distribution in the simulated sample is corrected to match the distribution in data. %

To account for a possible threshold enhancement in the $m(\lamp)$ spectrum, the partial branching fractions of the $\btlpmm$ decay are determined in two regions of $m(\lamp)$, below and above 2.8\gevcc, as done in Ref.~\cite{Belle:2007lbz}, according to

\begin{align}
{\cal B}_{\rm low(high)}(\btlpmm) & =\frac{N_{\rm low(high)}(\btlpmm)}{N(\btjpsilp)}\times \frac{\epsilon(\btjpsilp)}{\epsilon_{\rm low(high)}(\btlpmm)}\nonumber\\
& \times {\cal B}(\btjpsimmlp), %
\label{eq:bf}
\end{align}
where ${\cal B}(\btjpsimmlp) ={\cal B}(\btjpsilp)\times {\cal B}(\decay{\jpsi}{ \mumu})$, $N(\btjpsilp)$ and $\epsilon(\btjpsilp)$
 are the yields and efficiencies of the normalization mode, and  $N_{\rm low(high)}(\btlpmm)$ and $\epsilon_{\rm low(high)}(\btlpmm)$ are the corresponding parameters for the signal mode for the low (high) $m(\lamp)$ region.  The branching fractions related to the normalization mode are taken from Ref.~\cite{PDG2024}. 

 \begin{figure}[htb]
    \centering
    \includegraphics[width=0.32\textwidth]{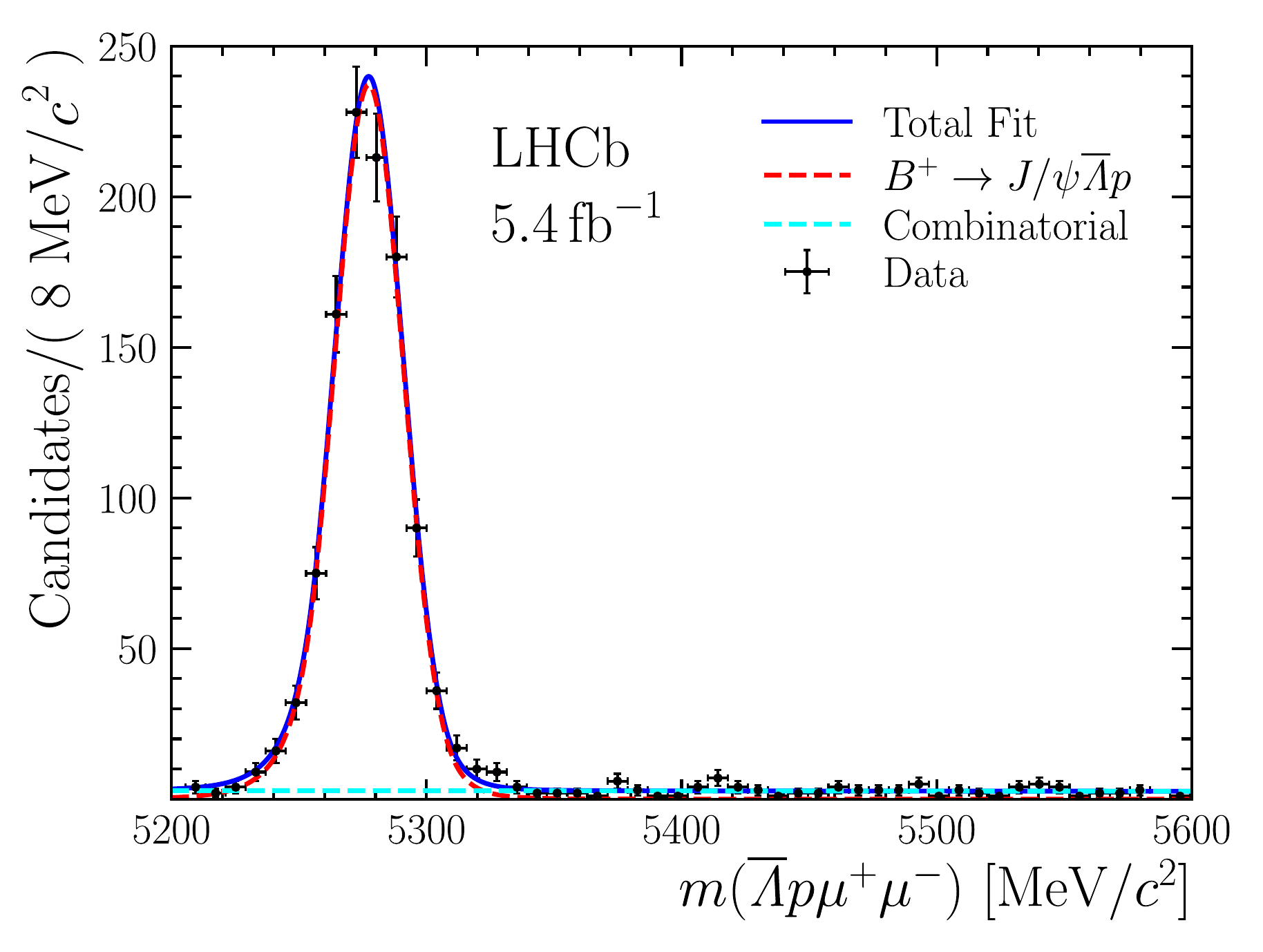}
    \includegraphics[width=0.32\textwidth]{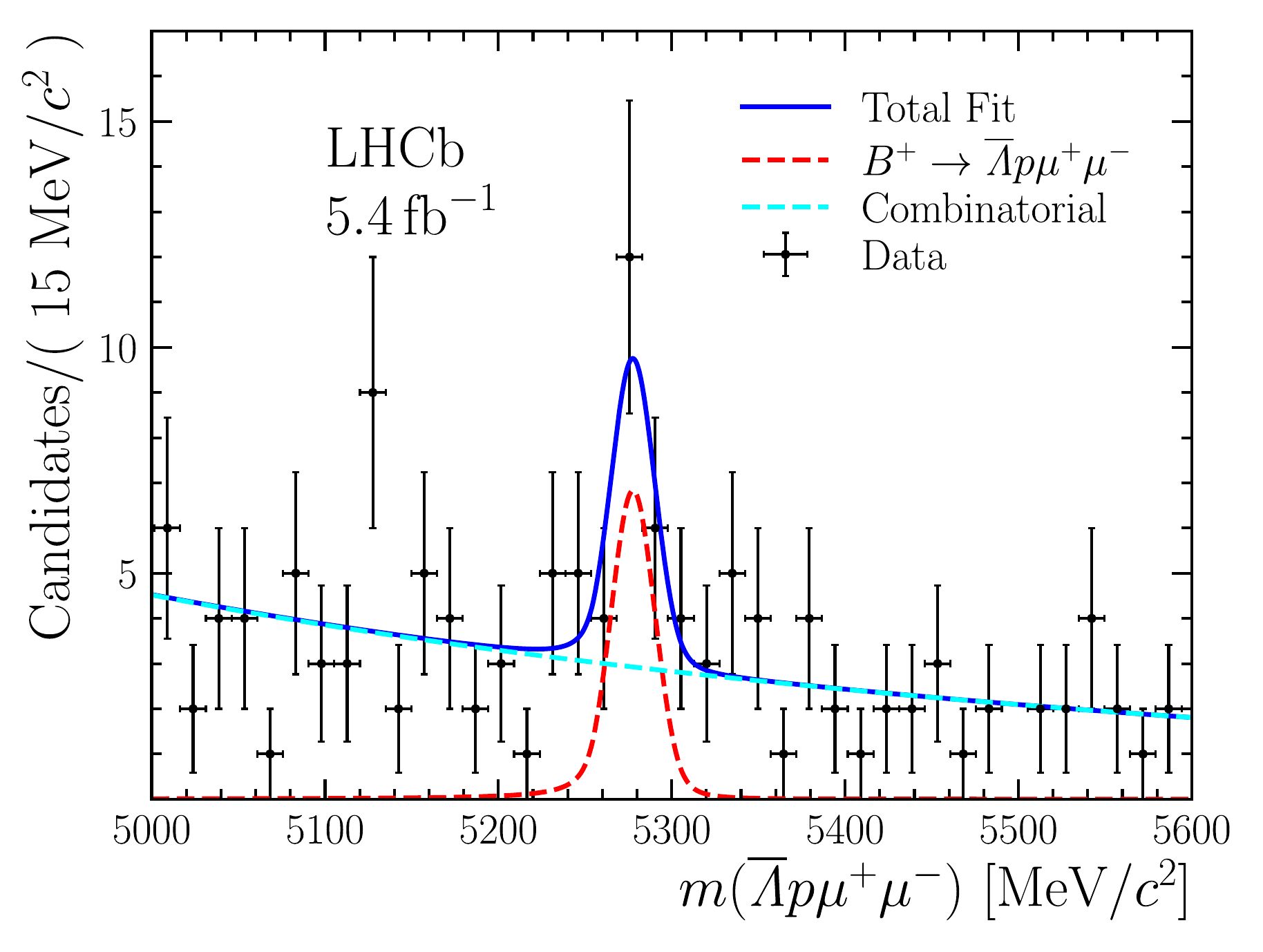}
    \includegraphics[width=0.32\textwidth]{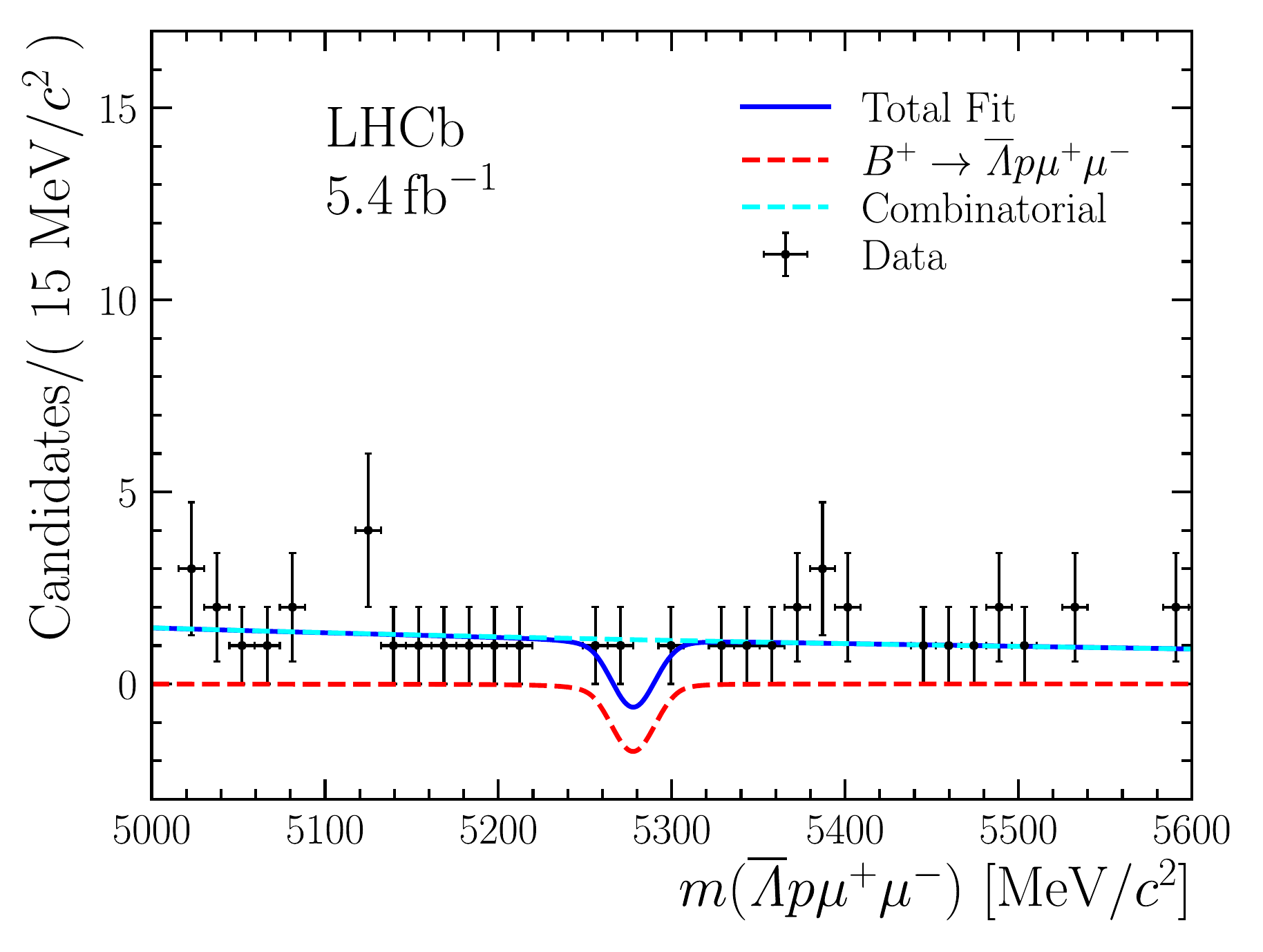}
    \caption{Invariant-mass distributions of  (left) $\btjpsilp$ and $\btlpmm$  candidates with (middle) $m(\lamp)<2.8\gevcc$ and (right)  $m(\lamp)>2.8\gevcc$, after applying all selection criteria described in the text. Candidates from the \emph{long} and \emph{downstream} $\lamb$ categories are combined for illustration purposes only. The combined fit functions are shown together with the individual fit components.}    
    \label{fig:SignalMasses}
\end{figure}

For the search for the rare \btlpmm decay, a simultaneous extended unbinned maximum-likelihood fit to the $m(\lamp \mumu)$ distributions is performed using the {\sc zfit}~\cite{py:zfit:2020} package. The fit is carried out separately in the \emph{long} and \emph{downstream} reconstruction categories and simultaneously includes both the signal and normalization modes. The resulting fit is used to determine the partial branching fractions in the two $m(\lamp)$ regions.
In the normalization mode, the $\btjpsilp$ decay is modeled using a double-sided Crystal Ball~(DSCB) function~\cite{Skwarnicki:1986xj}, while the combinatorial background is modeled using an exponential function with the slope parameter allowed to vary freely in the fit.
The DSCB shape parameters are fixed from simulation separately for the \emph{long} and \emph{downstream} categories, while one shift parameter $\delta_{\Bp}$ for the $\Bp$ peak position, and one scale parameter $\alpha_{\Bp}$ for the mass resolution, are included to account for differences between data and simulation. Both $\delta_{\Bp}$ and $\alpha_{\Bp}$ are allowed to vary freely in the fit. The $m(\lamp \mumu)$ distribution of the normalization mode for all data samples combined,  along with the fit projections, is shown in Fig.~\ref{fig:SignalMasses} (left) with high signal purity. The total yield of the normalization mode is determined to be $N(\btjpsilp)=1041\pm 34$.

The signal yields are parametrized using Eq.~(\ref{eq:bf}), and the partial branching fractions are shared between the samples from the same $m(\lamp)$ regions.  The model to describe the $m(\lamp \mumu)$ distribution is the same as that used for the $\btjpsilp$ normalization mode.  The model parameters for the signal component are fixed from simulation, except for $\delta_{\Bp}$ and $\alpha_{\Bp}$ that are shared with the normalization mode.

 \begin{figure}[tb]
    \centering
    \includegraphics[width=0.48\textwidth]{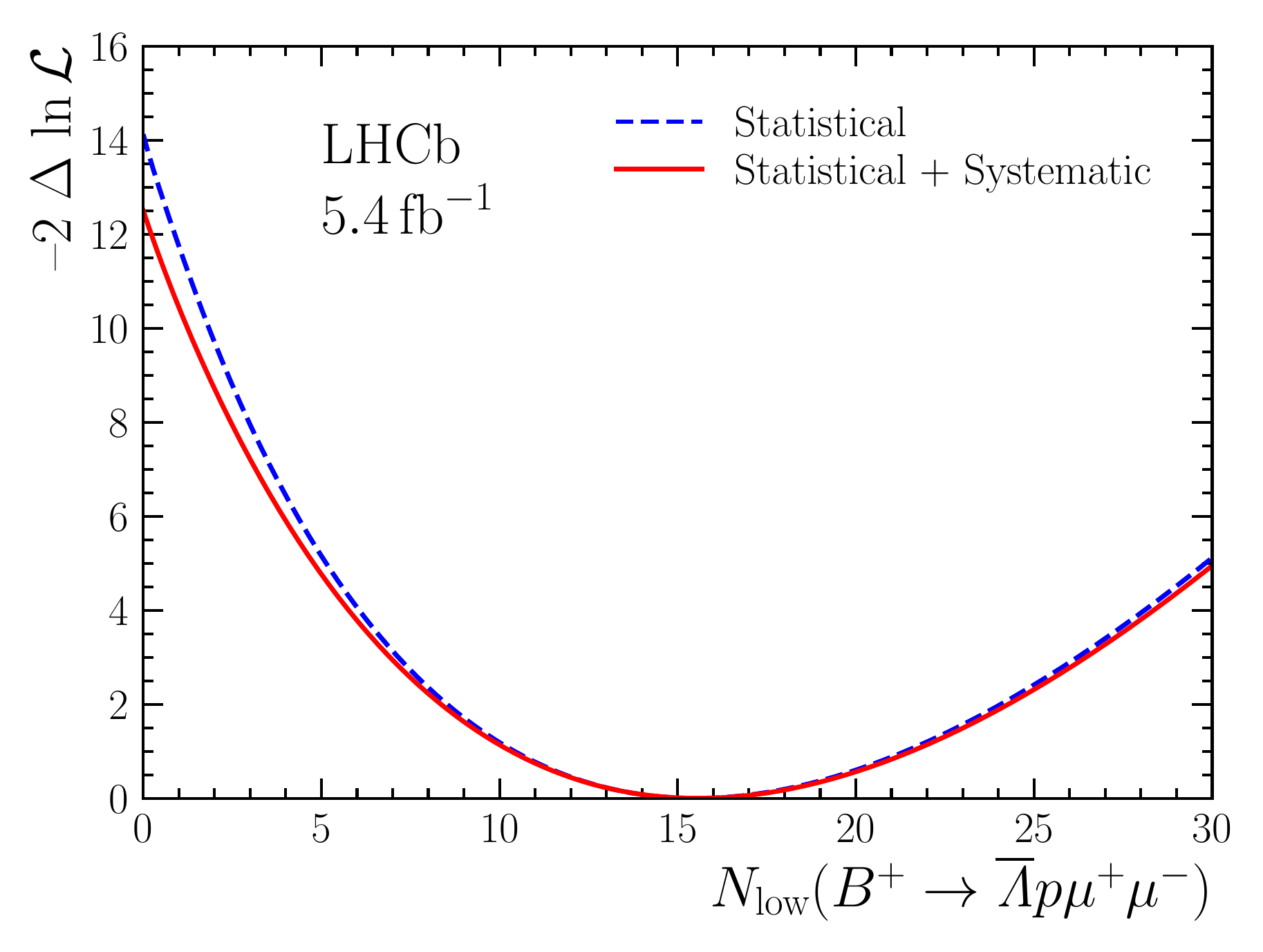}
    \caption{Projection of log-likelihood ratio $-2 \Delta \ln {\cal L}$ when determining the yield of $\btlpmm$ decays in the $m(\lamp)<2.8\gevcc$ region. }    
    \label{fig:LHscan}
\end{figure}

Contributions from misidentified backgrounds such as $\btlppipi$ and \decay{\Bp}{\Lambda_c^-(\to  \lamb \mu^{-}\bar\nu_\mu)p\pip} are found to be negligible and not considered in the fit. Figure~\ref{fig:SignalMasses} shows the $m(\lamp\mumu)$ 
distributions separately for the low and high $m(\lamp)$ regions, with fit projections also included. In the low $m(\lamp)$ region, a signal yield of 
$N_{\rm low}(\btlpmm)=15.5^{+5.9}_{-5.1}$
is obtained from the simultaneous fit to the data samples of the \emph{long} and \emph{downstream} categories. This corresponds to a statistical significance of 3.7$\sigma$, which is evaluated using
Wilks’s theorem~\cite{Wilks:1938dza}, through the difference in the log-likelihood, %
$\sqrt{-2\Delta \ln {\cal L}}$, 
where $\Delta \ln {\cal L}\equiv \ln({\cal L}_0/{\cal L}_S)$, and ${\cal L}_S$ (${\cal L}_0$) is the maximum likelihood of the baseline fit (fit with signal yield fixed to zero). The log-likelihood curve from the baseline fit is shown in Fig.~\ref{fig:LHscan}, along with the log-likelihood curve convolved with a Gaussian function to account for systematic uncertainties, %
which leads to a significance of $3.5\sigma$. 
In the high $m(\lamp)$ region, no signal is observed, as the fit converges towards a negative central value for $N_{\rm high}(\btlpmm)$. Therefore  an upper limit on ${\cal B}_{\rm high}$ using the ${\rm CL}_{\rm s}$ method is determined with the asymptotic approximation~\cite{Read:2002hq,Cowan:2010js}, with the details shown in the End Matter. The total branching fraction of \btlpmm  in the full $m(\lamp)$ region is also measured, as shown in the End Matter.

\begin{table}[t]
\centering
\caption{Relative systematic uncertainties on $\BF( \btlpmm )$, in percentage, in the two $m(\lamp)$ regions, as well as in the full range of $m(\lamp)$. The total systematic uncertainty is evaluated by adding all contributions in quadrature, assuming the sources are uncorrelated. 
}
\label{tab:summary_syst_BF}
\resizebox{1\columnwidth}{!}{
\begin{tabular}{l c cc}

\hline\hline
Systematic source   & $m(\lamp)<2.8\gevcc$ & $m(\lamp)>2.8\gevcc$ & Full region\\ \hline
Simulated sample size	    & $\phantom{1}0.8$      & $\phantom{1}0.9$ & $\phantom{1}0.6$\\
Kinematic correction        & $\phantom{1}2.1$      & $\phantom{1}2.2$ &$\phantom{1}2.2$ \\
Tracking efficiency 					& $\phantom{1}0.8$      & $\phantom{1}0.8$& $\phantom{1}0.8$\\ 
Trigger efficiency                     & $\phantom{1}2.0$      & $\phantom{1}4.4$&  $\phantom{1}3.4$\\ 
PID 						& $\phantom{1}2.8$      & $\phantom{1}2.9$ & $\phantom{1}2.9$ \\ 
Signal shape 				& $\phantom{1}7.0$      & $\phantom{1}7.0$ & $\phantom{1}6.2$  \\ 
Background shape 			& $\phantom{1}3.5$      & $\phantom{1}3.5$& $\phantom{1}3.8$ \\
Decay model correction (normalization) & $\phantom{1}2.6$ & $\phantom{1}2.6$ &  $\phantom{1}2.6$         \\
Decay model correction (signal)  & $\phantom{1}4.1$     & $\phantom{1}4.1$  &$\phantom{1}4.1$    \\ 
\hline
Total systematic 			&  $\phantom{1}10.1$    & $\phantom{1}10.8$ & $\phantom{1}10.1$ \\ 
\hline\hline%
\end{tabular} 
}
\end{table}

Different sources of systematic uncertainty affecting the branching fraction measurements are listed in Table~\ref{tab:summary_syst_BF}, and include those associated with the fit models, efficiencies, PID performance. %
Three sources of systematic uncertainties associated with the efficiencies estimated from simulation are determined. First, the uncertainties on the  efficiencies due to the size of the simulated samples are evaluated. Second, the uncertainty due to the weights used for correcting kinematic differences between simulation and data is assessed by 
varying the parameters that affect the performance of the weighting algorithm  individually. 
The total efficiencies are recalculated using the alternative sets
of weights, with the largest deviation taken as the associated systematic uncertainty.
Third, uncertainties on the tracking efficiencies, due to the detector material budget distribution and the particle interaction cross-sections, lead to uncertainties on the efficiency ratios. 
The hardware-trigger efficiencies are measured in bins of muon \pt using the \decay{\Bp}{\jpsi \Kp} data sample, and the related uncertainties are assigned by taking the variations in efficiencies with an alternative binning scheme.
Differences between PID efficiencies in data and simulation are corrected using a combination of simulated and data calibration samples. The variations in efficiencies after using an alternative PID correction method are taken as the systematic uncertainties. Details on the PID correction methods can be found in Ref.~\cite{LHCb-PAPER-2022-004}.

The systematic uncertainties on the signal and background shapes are evaluated by using alternative fit models. The signal shapes are varied in two ways: the DSCB function parameters are varied within their uncertainties, and an alternative Johnson $S_U$ function~\cite{JohnsonSU} is used to model the signal decay. For the signal decay, the background model is replaced by a linear function. Concerning the weights used to correct the $\btjpsilp$ decay model for the normalization channel, accounting for differences between data and simulation, 
the same method as for evaluating the uncertainty on the kinematic corrections is used to assign a systematic uncertainty.
The uncertainty related to the signal decay model is estimated by assuming that the $\lamp$ system originates exclusively from the $X(2085)$ resonance with the mass and width set to the  values reported in Ref.~\cite{BESIII:2023kgz}. Subsequently the signal decay of \decay{\Bp}{X(2085)\mumu} is generated with the model described in Ref.~\cite{Hatanaka:2008gu}  for the decay \decay{\Bp}{K_1(1270)^+\mumu}, and weighted to correct for the large mass of the $\lamp$ system. The resulting variation in signal efficiency is considered as the related systematic uncertainty, which also accounts for effects on the signal efficiency due to
the charmonium veto.
To assess the systematic uncertainties associated with a possible bias in the fit procedure, an ensemble of 
2\,000 pseudoexperiments is generated according to the baseline fit results. %
The uncertainties, considered as the mean biases observed when fitting these pseudoexperiments, are found to be negligible in all cases.

In summary, the first search for a semileptonic \decay{b}{ s\mumu} decay with a baryon-antibaryon pair in the final state is performed, using $pp$ collision data collected over the period 2016--2018 at $\sqrt{s}=13\tev$ and corresponding to an integrated luminosity of  5.4\invfb.
In the threshold-enhanced region of $m(\lamp)<2.8\gevcc$, the signal yield is found to be 
\mbox{$N_{\rm low}(\btlpmm) = 15.5 ^{+5.9}_{-5.1}$}
events, corresponding to a significance of $3.5\sigma$, after incorporating systematic uncertainties affecting the signal yield determination.
The branching fraction relative to the normalization channel is determined to be 
\mbox{$\BF_{\rm low}(\btlpmm)/\BF(\btjpsimmlp) = \left(1.95^{ +0.75}_{-0.64}(\rm stat)\pm 0.20(\rm syst)\right)$\%},  
from which the absolute branching fraction is determined to be 
\begin{equation*}
   \BF_{\rm low}(\btlpmm) = \left(1.70^{+0.65}_{-0.56}(\rm stat)\pm0.17(\rm syst)\pm0.14(\rm ext)\right) \times 10^{-8}, 
\end{equation*}
where the last uncertainty arises from the external branching fractions of the normalization channel. With no signal observed in the high $m(\lamp)$ region,  an upper limit at the 90\% (95\%) confidence level of 
\mbox{$\BF_{\rm high}(\btlpmm) < 2.8\,(3.7) \times 10^{-9}$}
is set. Integrated over the full available $m(\lamp)$ region, the total branching fraction is measured to be %
\mbox{$\BF(\btlpmm)=\left(1.53^{+0.69}_{-0.60}(\text{stat}) \pm 0.16(\text{syst}) \pm 0.13(\text{ext})\right) \times 10^{-8}$}.
Although lower than the prediction in Ref.~\cite{Dutta:2014yta}, the result is still compatible within 2$\sigma$. %
These measurements of $\BF(\btlpmm)$ hint at a manifestation of threshold enhancement in a \decay{b}{s\ell^+\ell^-} decay, and could lead to improved understanding of baryonic form factors in $B$-meson decays into a pair of light baryons. 
With the foundations laid by this analysis, the rich dynamics in the $\btlpmm$ decay are expected to motivate more detailed theoretical and experimental studies in the near future.

%% file: acknowledgements.tex
\section*{Acknowledgements}
%
%
\noindent We express our gratitude to our colleagues in the CERN
accelerator departments for the excellent performance of the LHC. We
thank the technical and administrative staff at the LHCb
institutes.
We acknowledge support from CERN and from the national agencies:
ARC (Australia);
CAPES, CNPq, FAPERJ and FINEP (Brazil); 
MOST and NSFC (China); 
CNRS/IN2P3 (France); 
BMFTR, DFG and MPG (Germany);
INFN (Italy); 
NWO (Netherlands); 
MNiSW and NCN (Poland); 
MCID/IFA (Romania); 
MICIU and AEI (Spain);
SNSF and SER (Switzerland); 
NASU (Ukraine); 
STFC (United Kingdom); 
DOE NP and NSF (USA).
We acknowledge the computing resources that are provided by ARDC (Australia), 
CBPF (Brazil),
CERN, 
IHEP and LZU (China),
IN2P3 (France), 
KIT and DESY (Germany), 
INFN (Italy), 
SURF (Netherlands),
Polish WLCG (Poland),
IFIN-HH (Romania), 
PIC (Spain), CSCS (Switzerland), 
and GridPP (United Kingdom).
We are indebted to the communities behind the multiple open-source
software packages on which we depend.
Individual groups or members have received support from
Key Research Program of Frontier Sciences of CAS, CAS PIFI, CAS CCEPP, 
Minciencias (Colombia);
EPLANET, Marie Sk\l{}odowska-Curie Actions, ERC and NextGenerationEU (European Union);
A*MIDEX, ANR, IPhU and Labex P2IO, and R\'{e}gion Auvergne-Rh\^{o}ne-Alpes (France);
Alexander-von-Humboldt Foundation (Germany);
ICSC (Italy); 
Severo Ochoa and Mar\'ia de Maeztu Units of Excellence, GVA, XuntaGal, GENCAT, InTalent-Inditex and Prog.~Atracci\'on Talento CM (Spain);
SRC (Sweden);
the Leverhulme Trust, the Royal Society and UKRI (United Kingdom).

%% file: endmatter.tex
\clearpage
\section*{End Matter}

\section{Branching fraction upper limit in the region \boldmath{\mbox{$m(\lamp)>2.8\gevcc$}}}
\label{Mat:ULhighM}

 \begin{figure}[htb]
    \centering
    \includegraphics[width=0.48\textwidth]{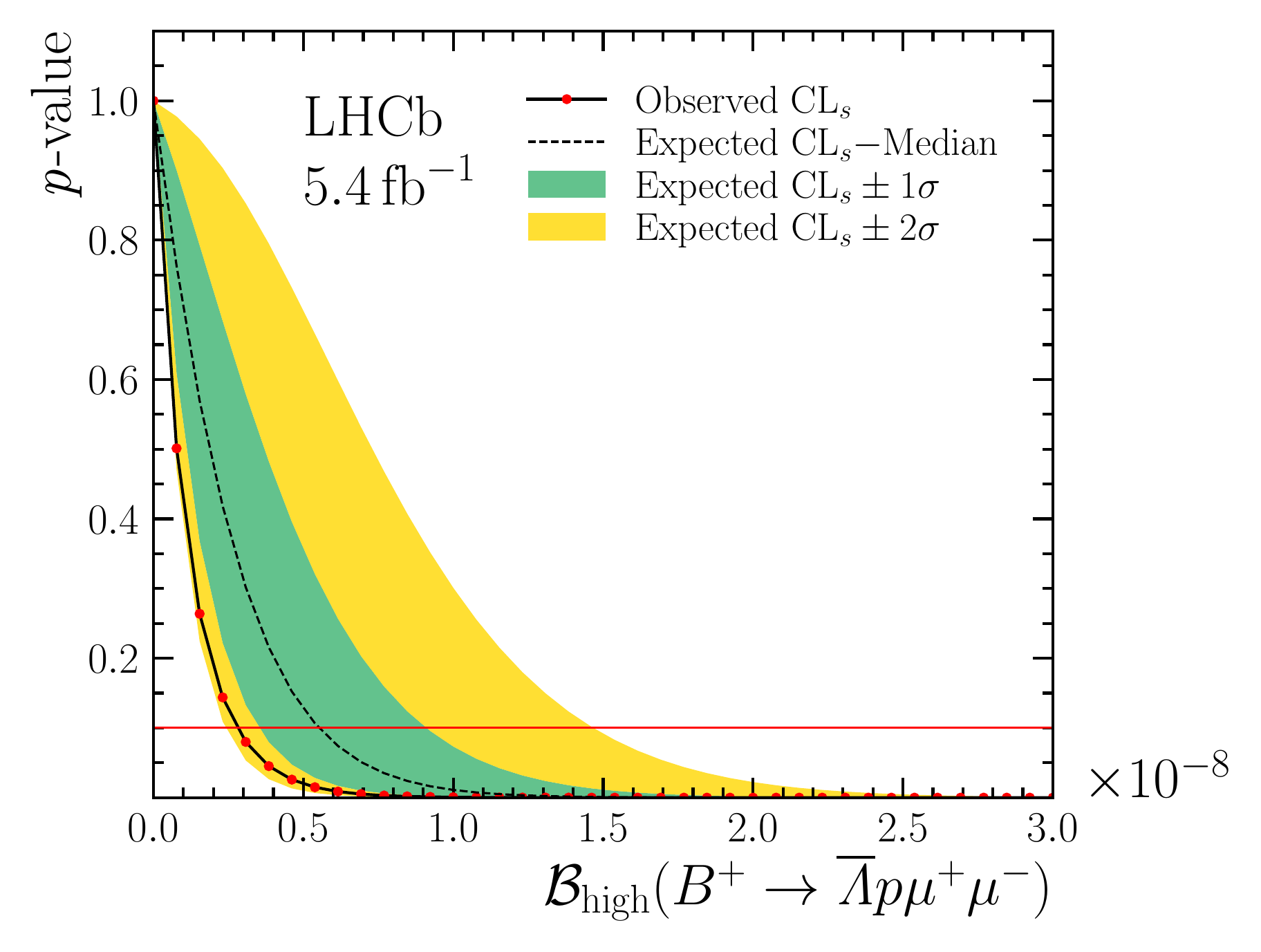}
    \caption{Results from the CL$_s$ scan used to obtain the upper limit on the partial branching fraction of $\btlpmm$ decay in the $m(\lamp)>2.8\gevcc$ region. The background-only expectation is shown by the dashed black line, with the 1$\sigma$ and 2$\sigma$ bands shown as green and yellow bands, respectively. The observed values are displayed as a solid black line with red markers, and the solid red line intersecting with the observation indicates the limit at the 90\% confidence level. }
    \label{fig:CLsHighM}
\end{figure}

The ${\rm CL_s}$ curve as a function of the partial branching fraction of $\btlpmm$ decay in the $m(\lamp)>2.8\gevcc$ region is shown in Fig.~\ref{fig:CLsHighM}, where the upper limit at the 90\% confidence level is depicted. In the limit calculation, the systematic uncertainties discussed in the main text, together with the uncertainty on ${\BF}(\btjpsimmlp)$~\cite{PDG2024}, %
are incorporated by  constraining the corresponding parameters in the likelihood fit using Gaussian priors. For each parameter, the mean is set to its central value, and the width is taken as the associated uncertainty.

\section{Fit result in the full \boldmath{$m(\lamp)$}  region}
\label{Mat:fitfull}

 \begin{figure}[htb]
    \centering
    \includegraphics[width=0.48\textwidth]{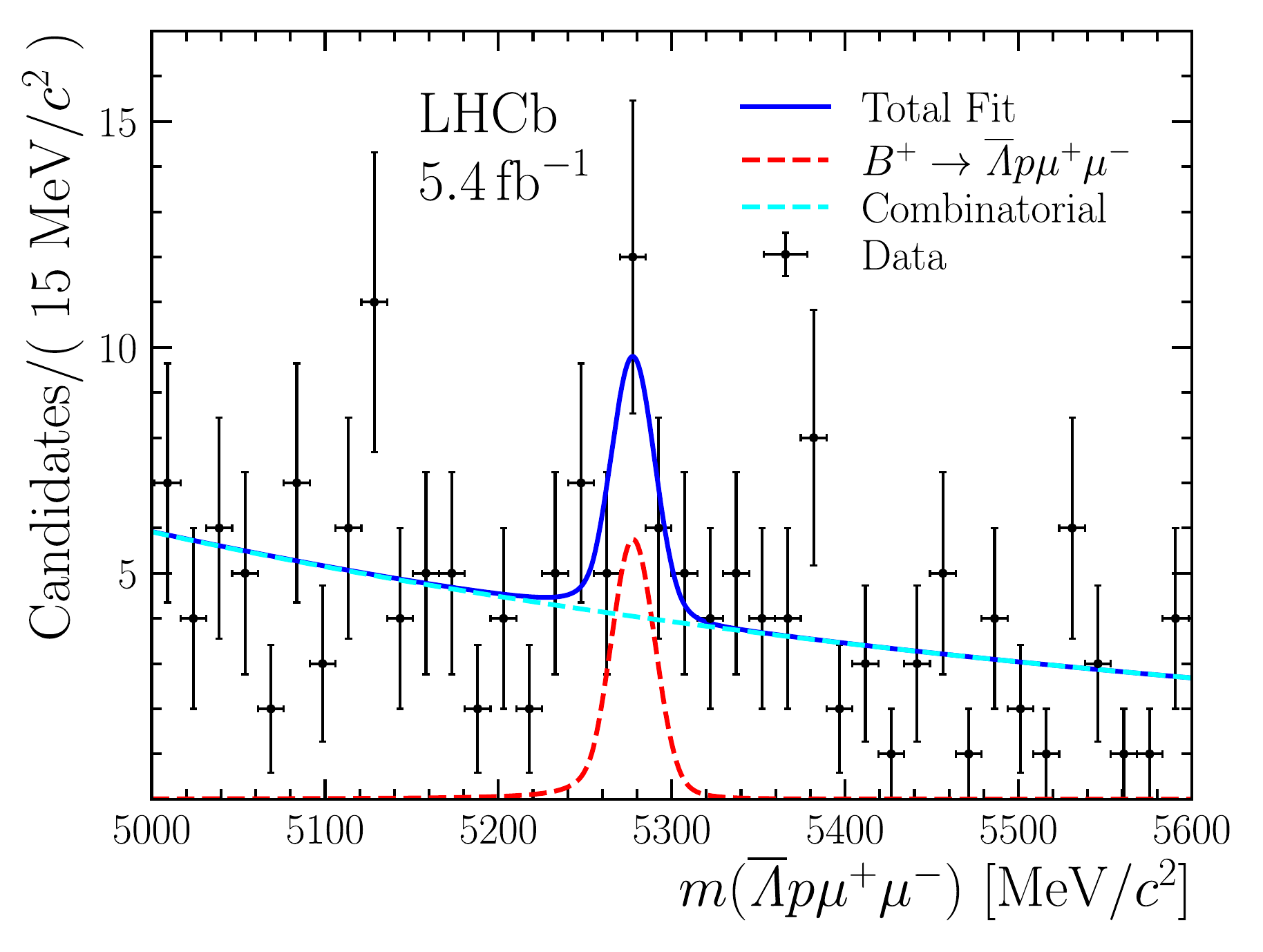}   
    \includegraphics[width=0.48\textwidth]{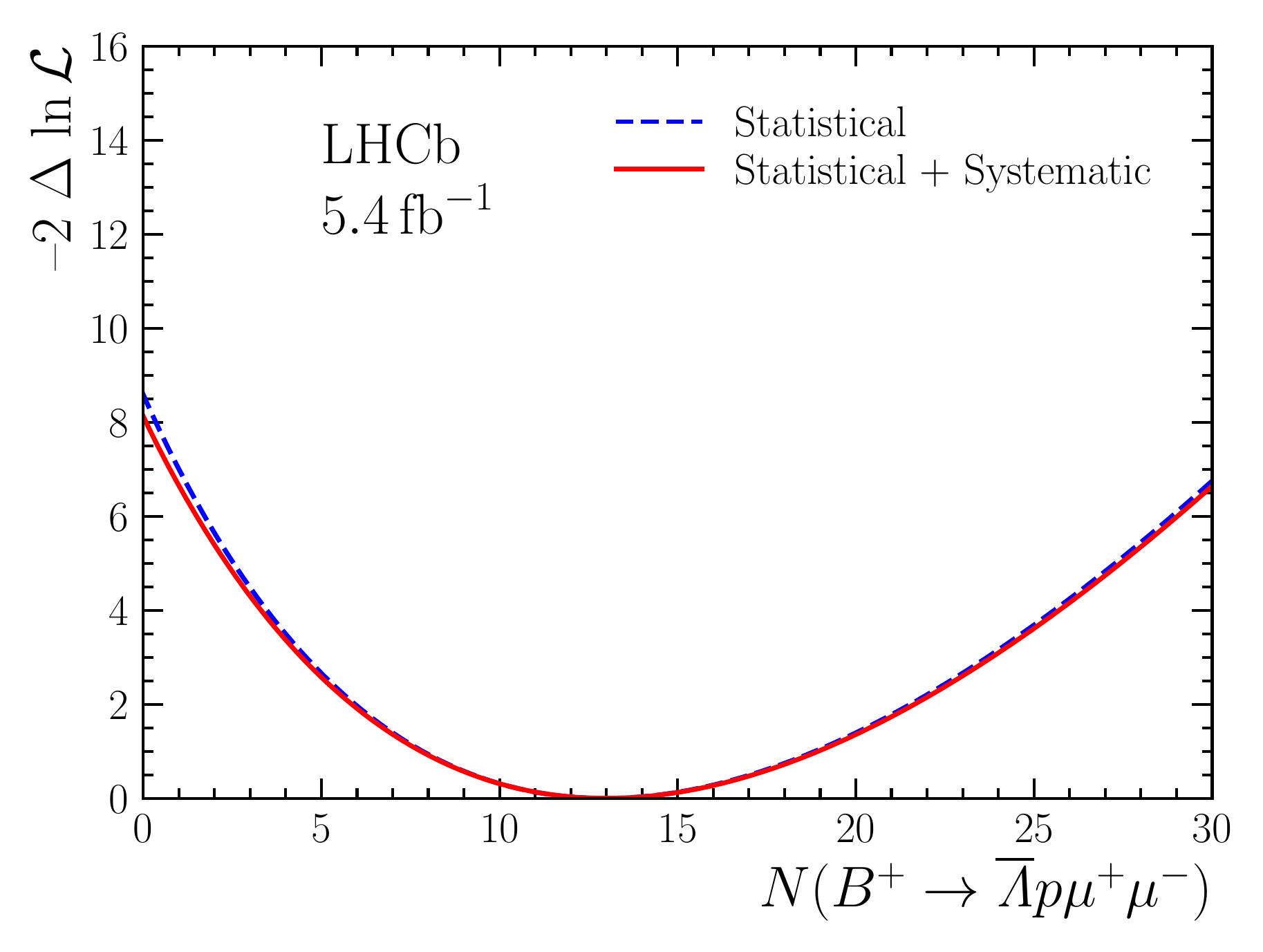}
    \caption{(Left) Invariant-mass distribution of $\btlpmm$  candidates, showing the fit result along with the fit model components. (Right) Projection of the likelihood ratio $-2 \Delta \ln {\cal L}$ used to determine the $\btlpmm$ yield in the full $m(\lamp)$ region.}
    \label{fig:FitFullReg}
\end{figure}

The two $m(\lamp)$ regions are combined to measure the total branching fraction of \mbox{$\btlpmm$}.
Figure~\ref{fig:FitFullReg} shows the $m(\lamp\mumu)$ distribution in the full $m(\lamp)$ region, with fit projection also included. A signal yield of $N(\btlpmm)=13.0^{+5.9}_{-5.1}$ is obtained, corresponding to a statistical significance of $2.9\sigma$. The likelihood curve from the baseline fit is also shown in Fig.~\ref{fig:FitFullReg}, along with the curve convolved with a Gaussian function to account for systematic uncertainties related to signal and background shapes, which slightly reduces the significance to 2.8$\sigma$. 

%% file: Authorship_LHCb-PAPER-2025-051.tex
\centerline
{\large\bf LHCb collaboration}
\begin
{flushleft}
\small
R.~Aaij$^{38}$\lhcborcid{0000-0003-0533-1952},
A.S.W.~Abdelmotteleb$^{58}$\lhcborcid{0000-0001-7905-0542},
C.~Abellan~Beteta$^{52}$\lhcborcid{0009-0009-0869-6798},
F.~Abudin\'en$^{58}$\lhcborcid{0000-0002-6737-3528},
T.~Ackernley$^{62}$\lhcborcid{0000-0002-5951-3498},
A.A.~Adefisoye$^{70}$\lhcborcid{0000-0003-2448-1550},
B.~Adeva$^{48}$\lhcborcid{0000-0001-9756-3712},
M.~Adinolfi$^{56}$\lhcborcid{0000-0002-1326-1264},
P.~Adlarson$^{86}$\lhcborcid{0000-0001-6280-3851},
C.~Agapopoulou$^{14}$\lhcborcid{0000-0002-2368-0147}
% ...

{\footnotesize \it

% ...
$^{14}$Universit{\'e} Paris-Saclay, CNRS/IN2P3, IJCLab, Orsay, France\\
% ...
$^{38}$Nikhef National Institute for Subatomic Physics, Amsterdam, Netherlands\\
% ...
$^{48}$Instituto Galego de F{\'\i}sica de Altas Enerx{\'\i}as (IGFAE), Universidade de Santiago de Compostela, Santiago de Compostela, Spain\\
% ...
$^{52}$Physik-Institut, Universit{\"a}t Z{\"u}rich, Z{\"u}rich, Switzerland\\
% ...
$^{56}$H.H. Wills Physics Laboratory, University of Bristol, Bristol, United Kingdom\\
% ...
$^{58}$Department of Physics, University of Warwick, Coventry, United Kingdom\\
% ...
$^{61}$School of Physics and Astronomy, University of Glasgow, Glasgow, United Kingdom\\
$^{62}$Oliver Lodge Laboratory, University of Liverpool, Liverpool, United Kingdom\\
% ...
$^{70}$Syracuse University, Syracuse, NY, United States\\
% ...
$^{86}$Department of Physics and Astronomy, Uppsala University, Uppsala, Sweden, associated to $^{61}$\\
% ...
}
\end{flushleft}